\documentclass{article}
\usepackage{graphicx}
\begin{document}

\date{}

\title{LA-UR-00-2629 \\COMPUTATION OF CHARGED-PARTICLE TRANSFER MAPS FOR GENERAL FIELDS AND GEOMETRIES USING ELECTROMAGNETIC BOUNDARY-VALUE DATA\\Paper presented at the 2000 International Computational Accelerator Physics Conference, Darmstadt, Germany, Sept.11-14, 2000}

\author{A.J. Dragt, P. Roberts, T.J. Stasevich, A. Bodoh-Creed \\
University of Maryland, College Park, Maryland 20742 USA\thanks{Work
supported in
part by DOE Grant DEFG0296ER40949.}
\\
\\
P. Walstrom \\ 
Los Alamos National Laboratory\\
Los Alamos, New Mexico 87545 USA}

\maketitle

\section{Introduction}
The passage of a charged particle through a region of
nonvanishing electromagnetic fields (e.g., a bending magnet, multipole
magnet,
spectrometer, electrostatic lens, electromagnetic velocity separator, etc.)
can be
described by a transfer map. In general, the map is a six-vector-valued
function
that relates the final six phase-space coordinates of a beam particle to
its initial
six phase-space coordinates. The map can be represented in either Taylor-
or
Lie-series form.  The series-expansion variables for the map are deviations
from a
nominal or reference trajectory, which in general is curved and must be
found by
numerical integration of the equations of motion of a reference particle.
The
reference particle is represented by a particular point in the initial
phase space
and a corresponding point in the final phase space.  Calculation of
aberration terms
(terms beyond lowest order) in the series form of the map requires
knowledge of
multiple derivatives of the electromagnetic fields along the reference
trajectory.

Three-dimensional field distributions associated with arbitrary realistic
beamline
elements can be obtained only by measurement or by numerical solution of
the
boundary-value problems for the electromagnetic fields. Any attempt to
differentiate
directly such field data multiple times is soon dominated by
``noise'' due to finite meshing and/or measurement errors. 

This problem can be overcome by the use of field data on a surface outside
of the
reference trajectory to reconstruct the fields along and around the
reference
trajectory. The present work is concerned with the static electric and
magnetic
fields in a source-free region inside of or near the beamline elements.
These
fields can be expressed as gradients of potential functions and are
harmonic
(solutions of Laplace's equation). The integral kernels for Laplace's
equation that
provide interior fields in terms of boundary data or boundary sources are
smoothing:
interior fields will be analytic even if the boundary data or source
distributions
fail to be differentiable or are even discontinuous. 

In our approach, we employ all three components of the field on the surface
to find a
superposition of single-layer and double-layer surface source distributions
that can
be used together with simple, surface-shape-independent kernels for
computing vector
potentials and their multiple derivatives (required for a Hamiltonian map
integration) at interior points. These distributions and kernels are found
by the
aid of Helmholtz's theorem. Kernels for derivatives are easily found by
differentiating the kernels for vector potentials with respect to the
field-point
variables. A novel application of the Dirac-monopole vector potential is
used to
find a kernel for the part of the vector potential that arises from the normal
component
of the field. 

These methods are the basis for map-generating modules that can be added to
 existing numerical electromagnetic field-solving codes and would produce transfer 
maps to any order for arbitrary static charged-particle beamline elements.

\section{Motion in a Static Magnetic Field}
The methods of this paper can be applied to both static electric and static 
magnetic fields, and combinations of the two.  For purposes of exposition, 
we will consider the case of magnetic fields.

In Cartesian coordinates and with the time $t$ as the independent variable,
the
Hamiltonian $H$ for motion of a particle of charge $q$ in a magnetic field
is given
by the relation
$$
H=[m^2c^4+c^2({\bf p}-q{\bf A})^2]^{1/2}\eqno (2.1)
$$ 
Here ${\bf A}$ is the vector potential associated with the {\bf B} field by
the
relation
$$
{\bf B}=\nabla \times {\bf A}. \eqno (2.2)
$$

For the purposes of generating maps it is more convenient to use one of the
coordinates, say the {\it z} coordinate, as the independent variable and to
treat the
time $t$ and its canonically conjugate momentum $p_t$ as dependent
variables$^1$. With
this choice of phase-space coordinates, the Hamiltonian $K$ for the motion
in a
magnetic field is given by the relation
$$
K=-[p_t^2/c^2-m^2c^2-(p_x-qA_x)^2-(p_y-qA_y)^2]^{1/2}-qA_z. \eqno (2.3)
$$
Typical equations of motion generated by this Hamiltonian are of the form
$$
x^{\prime}={{\partial K}\over{\partial p_x}} =
(p_x-qA_x)\bigl[p_t^2/c^2-m^2c^2-(p_x-qA_x)^2-(p_y-qA_y)^2\bigr]^{-1/2},
\eqno(2.4)
$$

\begin{eqnarray*}
p_x^{\prime}&=&-{{\partial K}\over{\partial x}}\\
&=&q\bigl[p_t^2/c^2-m^2c^2-(p_x-qA_x)^2-(p_y-qA_y)^2\bigr]^{-1/2}\Bigl[(p_x-q
A_x)
{{\partial {A_x}}\over {\partial{x}}}+(p_y-qA_y){{\partial{A_y}}\over
{\partial{x}}}\Bigr]\\&+&q{{\partial {A_z}}\over {\partial x}}.
\hspace{348pt} (2.5) 
\end {eqnarray*}

Evidently, the components of $\bf A$ and the first derivatives of the
components
of $\bf A$ with respect to $x$ and $y$ are required to compute
trajectories.  If we
take any one of these trajectories to be a reference trajectory, then
higher
derivatives of $\bf A$ with respect to $x$ and $y$ are also required to
compute the
transfer map $\cal M$ (first-order deviations and higher-order aberrations)
around
this trajectory. 

Suppose we write
$$
x=x^r+\xi , \eqno (2.6)
$$ 
$$
y=y^r+\eta , \eqno (2.7)
$$
where $x^r$ and $y^r$ refer to the reference trajectory. Then we require
Taylor
expansions of the form
$$
{\bf A}
(x^r+\xi,y^r+\eta,z)=\sum_\alpha{{\bf
A}^{\alpha}(x^r,y^r,z)P_{\alpha}(\xi,\eta)}
\eqno (2.8) 
$$
where the $P_{\alpha}$ are homogeneous polynomials labeled by some
convenient
index $\alpha$. Indeed, if we wish to compute an {\it l}th-order transfer
map, we
need to retain in the expansion (2.8) all homogeneous polynomials of degree
$l+1$ and lower.

\section{Modeling Field Data}
For general realistic magnetic beamline elements (e.g., iron-dominated
dipole
electromagnets, etc.), three-dimensional field data in the region about any
reference trajectory can presently be obtained only by measurement or by
numerical
solution of the boundary-value problem for the magnetic field. Data
specified at
discrete points on a three-dimensional grid can be smoothed and
interpolated by
various methods, but multiple differentiation of such field-data 
interpolants
introduces ``noise'' due to finite meshing and/or
measurement errors.  Moreover, the usual interpolating functions do not 
exactly satisfy  the Laplace equation, so their use is equivalent to introducing fictitious 
sources
into the source-free field volume.  In the two-dimensional case or in the case where
cylindrical
geometry can be employed, one may try to first smooth the data by fitting
it to some
assumed analytic form along some line (e.g. use some Enge profile) and then
extend
the form to points away from the line. In two dimensions (for example, in
fitting the
midplane fringe fields of a dipole with a straight field boundary) this can
be simply
done by analytically continuing the fitting function into the complex
plane. That is,
if the midplane field is $B_y(x)=F(x)$, where $F$ is the fitting function,
then
$B_y+iB_x=F(x+iy)$ for points away from the midplane.  The analogous
three-dimensional straight-axis approach in cylindrical coordinates
$(r,\phi,z)$ is
more complicated but qualitatively similar and is treated in Appendix A. 

Insight into the difficulties with this fitting approach can be found by
examining
the behavior of $F$ in the complex plane$^2$. In cases where the behavior
of $F(z)$
is dominated by poles, the poles represent effective line sources. If these
sources
do not approximate the distribution of the real sources of the physical
magnet, the
fit will diverge, frequently in a dramatic fashion, from the true field as
the field
evaluation point approaches the surfaces of the magnet. In the general
three-dimensional case any such fitting approach is even more
unsatisfactory.

The above difficulties can be eliminated by a method that uses field data
on some
surface that surrounds the reference trajectory and all nearby trajectories
of
interest in the beam-optics problem. In this approach, one exploits the
fact that the
integral kernels for Laplace's equation that provide interior fields in
terms of
boundary data or sources are smoothing: interior fields will be analytic
even if the
boundary data or sources fail to be differentiable or are even
discontinuous. 
Moreover, since harmonic functions assume their maxima on boundaries, a
method that
uses surface data is expected to be robust against errors.  That is,
interior
errors are expected to be no larger and generally smaller than errors in
the surface
data.

In classical potential theory, the Green's function for the Laplace
equation can be
used in principle to compute field values at arbitrary points inside a
surface that
encloses no sources. The field value at an arbitrary interior point can be
evaluated
by performing a surface integral of either the potential or the normal
component of the
field times the appropriate Green's function.  This approach is of no
practical use
for numerical computation in arbitrary geometries because the Green's
function for a
surface (which is a function of three field-point variables and two
surface-point
variables) is specific to that particular surface, varies from interior
point to
interior point, and cannot be expressed in closed or even series form,
except in the
case of certain special surfaces.

For the particular case of a straight or nearly straight reference
trajectory, we
have implemented two essentially equivalent approaches to expansions of the
vector
potential ${\bf A}$, making use of field data on a cylindrical surface that
have been
first analyzed into Fourier components in the azimuthal angle.  The field data
do not
have to have any particular symmetry; the only restriction is that the
sagitta of the
trajectory not be too large. Only one component of the field is needed; we use
the radial
component. The cylinder is chosen to be interior to any field windings,
iron, or
other magnetic sources. Both approaches employ a numerical Fourier integral
transform
with a modified Bessel-function kernel, and as expected, have been shown to
be robust
against errors. In the first approach$^3$, a subroutine recently added to
the MARYLIE
code$^4$ computes the on-axis gradient (see Appendix A for the definition of
on-axis
generalized gradient) and its $z$ derivatives from the surface data. The
routine
performs a numerical Fourier integral transform every time it is called
during
numerical map generation.  In the second approach$^5$, a separate program
uses a
different Fourier integral kernel to precompute a double-layer density
function of
$z$ (or stream function for equivalent currents) that is represented by its
values at
a collection of discrete $z$ points. These numbers are then read in by a
special
version of MARYLIE and used by a magnet-type subroutine that computes
gradients for
user-supplied stream functions.

The analytic methods described above could be implemented because the
cylinder is a
surface of constant radius in cylindrical coordinates, which represent one
of the
classic coordinate systems in which the Laplace equation is separable.  An
analogous
analytic approach can be found for any surface that can be described by
holding
constant one of the coordinates of a system in which the Laplace equation
is
separable. For example, in the case of a rectangular box$^6$, one can fit
double
trigonometric/hyberbolic function series with adjustable coefficients to
surface data
to find interior fields. 

However, for general geometries, the solution to Laplace's equation is not
available
in analytic form, and the methods available for separable-coordinate
geometries
cannot be used. In the next sections we describe a method that bypasses the
need for
an analytic solution to Laplace's equation for general geometries by making
use of
all three components of the surface field, and still provides the resultant
smoothing that arises from the use of surface data. Thus, this method too
is
expected to be robust against errors. As with some methods previously
described$^{7,8}$, it is based on use of effective sources on a surface
surrounding
the field-evaluation volume to represent interior fields. However, unlike
the method
of Ref. 8, the method we describe in the following section is not based on
the use
of least-square fits to find the source strengths. Instead, the sources are
found
directly from field data by use of Helmholtz's theorem.

\section{Helmholtz's Theorem}
Let ${\bf F}({\bf r})$ be any vector field and let $V^\prime$ be any volume
bounded by a surface $S^\prime$. Then, according to Helmholtz's theorem$^9$,
${\bf F}$
for ${\bf r}$ within $V^\prime$ can be written in the form,
$$
{\bf F}=\nabla\times{\bf A}+\nabla\Phi, \eqno (4.1)
$$
where ${\bf A}$ and $\Phi$ are defined by the integrals, 
$$
{\bf A}({\bf r})=-{1\over{4\pi}}\int_{S^\prime} {{{\bf n}({\bf
r}^\prime)\times{\bf
F}({\bf r}^\prime)}\over{| {\bf r}- {\bf r}^\prime |}}
dS^\prime+{1\over{4\pi}}\int_{V^\prime} {{\nabla^\prime\times{\bf F}({\bf
r}^\prime)}\over{| {\bf r}- {\bf r}^\prime | }}dV^\prime, \eqno (4.2)
$$

$$
{\Phi}({\bf r})={1\over{4\pi}}\int_{S^\prime} {{{\bf n}({\bf
r}^\prime)\cdot{\bf F}({\bf r}^\prime)}\over{| {\bf r}- {\bf r}^\prime |
}}dS^\prime-{1\over{4\pi}}\int_{V^\prime} {{\nabla^\prime\cdot{\bf
F}({\bf
r}^\prime)}\over{| {\bf r}- {\bf r}^\prime | }}dV^\prime. \eqno (4.3)
$$
Here, as usual, $\nabla$ denotes partial differentiation with respect to
the
components of ${\bf r}$, and $\nabla^{\prime}$ denotes partial
differentiation
with respect to the components of ${\bf r}^{\prime}$. Also, ${\bf n}({\bf
r}^{\prime})$ denotes the outward normal to $S^{\prime}$ at the point ${\bf
r}^{\prime}$. In the case that ${\bf F}$ is the magnetic field ${\bf B}$,
and under
the further assumption that ${\bf B}$ is curl and divergence free in
$V^{\prime}$
(i.e., the field is static and there are no sources in $V^{\prime}$), which
will be
true for our applications, ${\bf A}$ and $\Phi$ are given by surface
integrals alone
and (4.1) through (4.3) take the simpler form 
$$
{\bf B}=\nabla\times{\bf A}^t+\nabla\Phi, \eqno (4.4)
$$
with
$$
{\bf A}^t({\bf r})=-{1\over{4\pi}}\int_{S^\prime} {{{\bf n}({\bf
r}^\prime)\times{\bf
B}({\bf r}^\prime)}\over{| {\bf r}- {\bf r}^\prime | }}dS^\prime, \eqno
(4.5)
$$
$$
{ \Phi}({\bf r})={1\over{4\pi}}\int_{S^\prime} {{{\bf n}({\bf r}^\prime)
\cdot{\bf B}({\bf r}^\prime)}\over{| {\bf r}- {\bf r}^\prime | }}dS^\prime.
\eqno
(4.6)
$$
The superscript $t$ on ${\bf A}^t$ is used to indicate that ${\bf A}^t$
depends
only on the tangential components of ${\bf B}$ on the surface.

We have obtained interior fields in terms of surface fields. However, there
is one
defect. To employ canonical equations of motion using (2.3) we need to
obtain ${\bf
B}$ entirely from a vector potential as in (2.2) rather than from a sum of
vector and
scalar potentials as in (4.4). The next section describes how this defect
can be
overcome with the artifice of Dirac monopoles.

\section{Dirac Monopole Representation}
Let ${\bf G}^n$ denote the Dirac monopole vector potential$^{10,11}$ given by the relation
$$
{\bf G}^n({\bf r};{\bf r}^{\prime},{\bf m})={{{\bf m} \times
{({\bf r}-{\bf r}^{\prime})}} \over {4\pi}[|{\bf r}-{\bf r}^\prime|-{\bf m}
\cdot({\bf r}-{\bf r}^\prime)]|{\bf r}-{\bf r}^\prime|}.\eqno (5.1)
$$
The unit vector ${\bf m}$ in (5.1) points in the direction of the Dirac
string,
which is taken to be a straight line that extends from the point ${\bf
r}^\prime$
to infinity in the direction of ${\bf m}$. The vector ${\bf r}$ is the
field-evaluation point and ${\bf r}^\prime$ the source point. Eq. (5.1) is
derived
from Eq. 6.161 of Ref. 9 by explicitly evaluating the integral from zero to
positive infinity along the string. The vector field ${\bf G}^n$ is analytic in
${\bf r}$
except along the Dirac string. It has the desired property
$$
{{1}\over{4\pi}}\nabla{{1}\over{|{\bf r}-{\bf r}^\prime|}} =
\nabla\times{\bf G}^n
({\bf r};{\bf r}^\prime {\bf m}) \eqno (5.2)
$$
for all points ${\bf r}$ except those on the string. 

We now define a vector field ${\bf A}^n$ by the integral
$$
{\bf A}^n({\bf r})=\int_{S^\prime}[{\bf n} ({\bf
r}^\prime)\cdot{\bf B}({\bf r}^\prime)]{\bf G}^n[{\bf r};{\bf r}^\prime
{\bf n}({\bf
r}^\prime)]dS^\prime .
\eqno (5.3)
$$
In writing (5.3) we have taken the string direction ${\bf m}$ at each point
${\bf
r}^\prime$ in $S^\prime$ to lie along the outward normal ${\bf n}({\bf
r}^\prime)$.
(Other string direction choices are also possible, and in fact necessary
for certain
surfaces, such as toroids. Different choices of string directions simply
amount to
gauge transformations on ${\bf A}^n$.  The essential requirement is only
that the
strings do not intersect the volume of interest $V^\prime$.) Here we have
used the
notation ${\bf A}^n$ to indicate that ${\bf A}^n$ depends only on the
normal
component of ${\bf B}$. 

In view of (4.3). (5.1), and (5.2), ${\bf A}^n$ has the property
$$
\nabla\times{\bf A}^n=\nabla\Phi .\eqno (5.4)
$$
Therefore we may define a net vector potential ${\bf A}$ in terms of ${\bf
A}^n$
and ${\bf A}^t$ by the rule
$$
{\bf A}({\bf r})={\bf A}^n({\bf r})+{\bf A}^t({\bf r}),\eqno (5.5)
$$
with the result, in view of (4.4) and (5.4), that (2.2) holds for the ${\bf
A}$
given by (5.5), i.e., the total magnetic field inside $V^\prime$ is given
by the curl
of the vector potential of (5.5). The relations (4.5) and (5.5),
which give ${\bf A}^t$ and ${\bf A}^n$ entirely in terms of surface data,
have the
virtue that they can be differentiated repeatedly at will with respect to
the
components of ${\bf r}$ provided that ${\bf r}$ is within $V^\prime$. 
Indeed, they
show that ${\bf A}^t$ and ${\bf A}^n$, and hence ${\bf A}$, are (real)
analytic in
the components of ${\bf r}$ for ${\bf r}$ within $V^\prime$. 
Correspondingly, the
series (2.8) will converge in a finite domain in $V^\prime$ whose exact
size and
shape can be determined from the theory of functions of many complex
variables$^{12}$.

\section{Further Manipulations}
In view of (5.2) the vector field ${\bf G}^n$ has the property
$$
\nabla\times(\nabla\times{\bf G}^n)=0\eqno(6.1)
$$
for points ${\bf r}$ not on the string. It can be verified by direct
calculation
that ${\bf G}^n$ also satisfies the Coulomb gauge condition,
$$
\nabla\cdot{\bf G}^n = 0\eqno(6.2)
$$
for points ${\bf r}$ not on the string. It follows from (5.3) that ${\bf
A}^n$
also has these properties,
$$
\nabla\times(\nabla\times{\bf A}^n)=0,\eqno(6.3)
$$
$$
\nabla\cdot{\bf A}^n=0, \eqno(6.4)
$$
for all points ${\bf r}$ in $V^\prime $. Eq. 6.3 means that the field computed from
${\bf
A}^n$ is curl-free; by Maxwell's equations a non-zero curl would imply that
error
currents are present inside $V^\prime$. We note that both these relations
hold no
matter what the factor ${\bf n}({\bf r}^\prime) \cdot {\bf B}({\bf
r}^\prime)$ in
(5.3) is and no matter how badly the integral (5.3) is evaluated (say by
numerical
methods) since these relations depend only on the underlying properties
(6.1) and
(6.2) of the kernel ${\bf G}^n$. All that is required is that the kernel
${\bf G}^n$
be evaluated properly.

We would like to have analogous properties for ${\bf A}^t$, the part of the
vector
potential due to the components of ${\bf B}$ that are tangent to the
surface. We
first note that the kernel $1/|{\bf r}-{\bf r}^\prime|$ satisfies the
Laplace
equation, i.e., 
$$
\nabla^2 {{1}\over{|{\bf r}-{\bf r}^\prime|}} = 0. \eqno (6.5)
$$
From this relation and the definition (4.5) it follows that
$$
\nabla^2{\bf A}^t = 0\eqno (6.6)
$$
for all points ${\bf r}$ in $V^\prime$ no matter what the vector function
${\bf
n}({\bf r}^\prime) \times{\bf B}({\bf r}^\prime)$ in (4.5) is and no matter
how
poorly the integral is evaluated. For (6.6) to be satisfied, all that is
required is
that the kernel $1/|{\bf r}-{\bf r}^\prime|$ be evaluated properly. Recall
also the
vector identity
$$
\nabla\times(\nabla\times{\bf A}^t) = -\nabla^2{\bf A}^t +
\nabla(\nabla\cdot{\bf
A}^t). \eqno (6.7)
$$
We see from (6.6) and (6.7) that ${\bf A}^t$ will satisfy the condition
$\nabla
\times (\nabla\times{\bf A}^t)=0$, i.e., the part of ${\bf B}$ coming from
${\bf
A}^t$ will have zero curl, if ${\bf A}^t$ satisfies the condition
$\nabla\cdot{\bf
A}^t=0$. 

As it stands, it can be shown that (5.4) holds for ${\bf A}^t$ providing
the
whole integral (4.5) including its full integrand are evaluated properly.
What we
would like to do is transform the integrand in such a way that relations
analogous to
both (6.3) and (6.4) will hold for ${\bf A}^t$ no matter how badly the
integral and
parts of its integrand are evaluated, provided that only a certain kernel
is
evaluated properly. This is easily done with the aid of a scalar potential.
Since
${\bf B}$ is curl-free within $V^\prime$ and on the surface $S^\prime$, we
know
that there is a scalar potential $W$ with the property
$$
{\bf B}=-\nabla^\prime W({\bf r}^\prime)\eqno (6.8)
$$
for ${\bf r}^\prime$ on $S^\prime $ . Using the integral identity
$$
\int_{S^\prime}{{{\bf n}({\bf r}^\prime)\times\nabla^\prime W({\bf
r}^\prime)}\over{|{\bf r}-{\bf r}^\prime|}}dS^\prime=\int_{S^\prime}W({\bf r}^\prime
){\bf n}({\bf r}^\prime)\times\nabla^\prime {{1}\over{|{\bf r}-{\bf r}^\prime|}}dS^\prime \eqno(6.9)
$$
that holds for any function $W$, the integral (4.5) for ${\bf A}^t$ can be
rewritten in the form
$$
{\bf A}^t({\bf r})=\int_{S^\prime}W({\bf r}^\prime){\bf G}^t[{\bf r};{\bf r}^\prime,{\bf n}({\bf r}^\prime)]dS^\prime,\eqno
(6.10)
$$
where ${\bf G}^t$ is the kernel given by the expression
$$
{\bf
G}^t[{\bf r};{\bf r}^\prime,{\bf n}({\bf r}^\prime)]={{1}\over{4\pi}}{\bf n}({\bf r}^\prime)\times\nabla^\prime{{1}\over{|{\bf r}-{\bf r}^\prime|}}.
\eqno
(6.11)
$$
Note that in (6.10) the only values of $W$ that are required to compute 
${\bf A}^t$ are those
on the surface ${\bf S}^\prime$. We also note from (6.8) that, up to an inconsequential
constant, the values of $W$ at points ${\bf r}$ in ${\bf S}^\prime$ are completely
specified by
knowing the tangential components of ${\bf B}$ on ${\bf S}^\prime$. We further note that
(6.11)
is the vector potential of an infinitesimal dipole at ${\bf r}^\prime$ with magnet
moment vector
normal to the surface; $W$ can therefore be identified with a double-layer
density
distribution on the surface. 

It can be verified that the kernel ${\bf G}^t$ given by
(6.11) has the properties
$$
\nabla^2{\bf G}^t=0,\eqno(6.12)
$$
$$
\nabla\cdot{\bf G}^t=0.\eqno(6.13)
$$
It follows that, when the representation (6.10) is used, the vector field ${\bf A}^t$ 
satisfies curl and divergence relations analogous to those of (6.3) and
(6.4),
i.e., it satisfies the relations
$$
\nabla\times(\nabla\times{\bf A}^t)=0\eqno(6.14)
$$
$$
\nabla\cdot{\bf A}^t=0\eqno(6.15)
$$
for ${\bf r}$ in $V^\prime$ no matter how badly the integral (6.10) is evaluated
and no
matter what values are used for $W$ on $S^\prime $. All that is required is that
the kernel
${\bf G}^t$ be evaluated properly on $S^\prime $. 

As a consequence of (5.5), (6.3), (6.4), (6.14), and (6.15) we are
guaranteed that
for the total vector potential and field, respectively,
$$
\nabla\cdot{\bf A}=0,\eqno(6.16)
$$
and
$$
\nabla\times{\bf B}=\nabla\times(\nabla\times{\bf A})=0 \eqno (6.17)
$$
for ${\bf r}$ within $V^\prime$ no matter how badly the integrals (5.3) and (6.10)
are
evaluated and no matter what values are used for ${\bf n}({\bf r}^\prime)\cdot{\bf B}({\bf r}^\prime)$
and
$W({\bf r}^\prime)$ for ${\bf r}^\prime$ in $S^\prime $. Under the same conditions the vector potential
${\bf A}({\bf r})$ will also be analytic for ${\bf r}$ in $V^\prime$. Again,
all that
is required is that the kernels ${\bf G}^n$ and ${\bf G}^t$ be evaluated
properly.
Finally, since the divergence of a curl always vanishes, use of (2.2)
guarantees that
$$
\nabla\cdot{\bf B}=\nabla\cdot(\nabla\times{\bf A})=0 \eqno (6.18)
$$
will always hold.  

Of course, the better we evaluate the integrals (5.3) and (6.10) including
their
integrands, the better the ${\bf B}$ given by (2.2) with the use of (5.3),
(5.5),
and (6.10) will agree with the true ${\bf B}$. However, no matter what
(provided the
kernels ${\bf G}^n$ and ${\bf G}^t$ evaluated properly), the physically
required
conditions (6.16) and (6.17) will be met.

In the language of classical potential theory, we have found a combination
of a
single-layer and a double-layer (dipole) distribution on the surface and an
associated pair of vector-valued integral kernels that together produce a
total
vector potential that, in turn, gives a curl-free magnetic field inside
$V^\prime$
that replicates the field of the beamline element. The single layer arises
from
the normal component of the field at the surface and the double layer from
the
transverse field components, or equivalently, from the scalar potential at
the
surface.  Indeed, the two density distributions could have been found by
use of
Green's theorem$^{13}$ without explicit use of the Helmholtz theorem.

\section{Implementation}
The kernels ${\bf G}^n$ and ${\bf G}^t$ given by (5.1) and (6.11) may be
expanded
analytically, say by some symbolic manipulation program such as
Mathematica, to give
expressions of the form
$$
{\bf G}^n(x^r+\xi,y^r+\eta,z^r;{\bf r}^\prime,{\bf m}) = \sum_\alpha {\bf
G}^{n\alpha}
(x^r,y^r,z^r;{\bf r}^\prime,{\bf m})P_\alpha (\xi,\eta),\eqno(7.1)
$$ 
$$
{\bf G}^t(x^r+\xi,y^r+\eta,z^r;{\bf r}^\prime,{\bf m})=\sum_\alpha {\bf G}^{t\alpha}
(x^r,y^r,z^r;{\bf r}^\prime,{\bf m})P_\alpha (\xi,\eta),\eqno(7.2)
$$
These expansions can then be inserted into (5.3) and (6.10) and the
surface
integrals over $S^\prime$ performed numerically to yield the desired
expansion (2.8). Since the expansions (7.1) and (7.2) have been carried out
analytically, the expansion (2.8) will be consistent with the conditions
(6.16) and
(6.17) even if the surface integrals are not done perfectly. 

The program just described has been implemented for the case of a bent
rectangular
box with straight end arms as illustrated in Figure 6.1 below (see Appendix
B for a more
detailed discussion of possible surface topologies, i.e., boxes, tori,
etc., and
associated choices of the orientation of the Dirac string). The volume 
$V^\prime$ enclosed
by such a box is well suited to integrating reference trajectories and
finding maps
about these trajectories for the case of bending magnets. 

Preliminary results have been obtained for dipole-like fields that can be
modeled
analytically (e.g. fields produced by a superposition of magnetic monopoles
located
outside $V^\prime $). For these model fields surface normal fields and surface
scalar
potentials on $S^\prime$  can be calculated, and reference trajectories and their
associated
transfer maps can then be calculated using the expansion (2.8) obtained by
integrating over $S^\prime $ . For these model fields, reference trajectories and
transfer
maps about them can also be obtained from the direct analytic expansion of
the
associated model vector potential.

Numerical comparisons were made for fields computed in three ways: 1. directly
from the sources outside of the surface, 2. by integrating the kernels
(5.1) and
(6.11) times source strengths over the surface, and 3. from expansions
(like that of
Eq. 2.8, but in three variables) obtained by integrating the derivatives of
the
kernels times source distributions over the surface. It was found that the
vector
potential and its derivatives computed by the three methods agreed well for
points
within $V^\prime$. Correspondingly, it was found that the reference
trajectories and their associated maps obtained by the two methods agreed
well.  

For problems of actual interest the surface normal fields and surface
scalar
potentials on $S^\prime$ will be determined numerically by the use of some
3-dimensional field code or by direct measurement, and the surface
integrals (5.3)
and (6.10) over the expansions (7.1) and (7.2) will be carried out using
these
numerical values. Based on the test results obtained so far, there is every
reason
to believe that reference trajectories and their associated transfer maps
computed
from these surface data should be just as highly accurate as it was for the
test
cases. That is, by using the methods of this paper, modules can be added to
existing
electromagnetic codes that will produce reliably, when requested,
associated
transfer maps to any order for arbitrary static charged-particle beamline
elements.
Thus it should now be possible, for the first time, to design and analyze
the effect
of general static beam-line elements, including all fringe-field and error
effects,
in complete detail.
\vspace{0.3cm}
{\par\centering \includegraphics{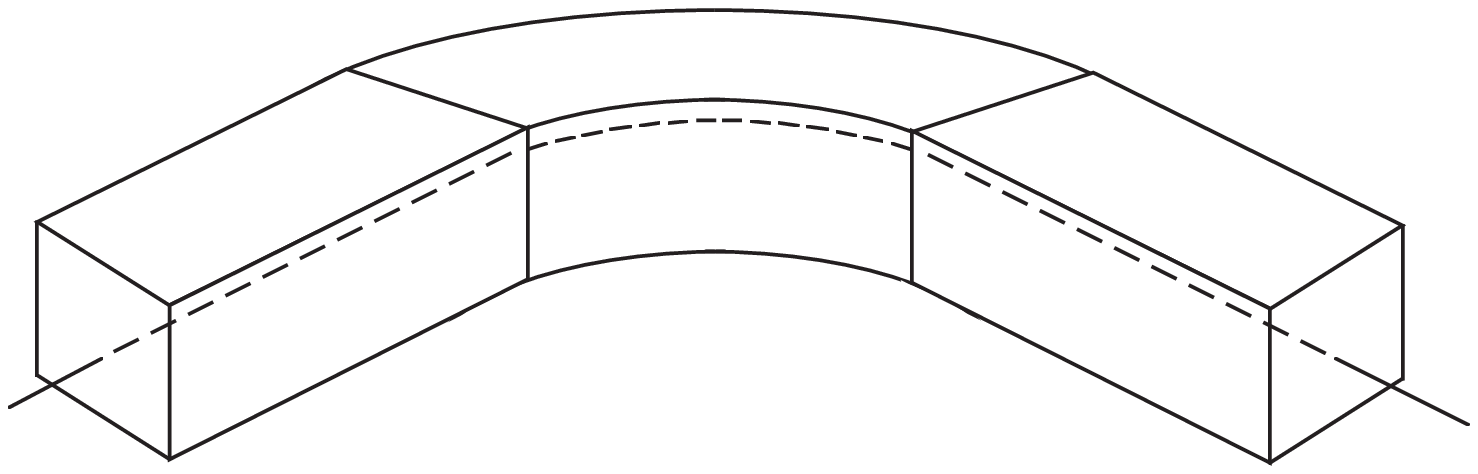} \par}
\vspace{0.3cm}

\parindent=0in
Figure 6.1. A volume $V^\prime$ consisting of a bent rectangular box with
straight arms suitable for treating a bending magnet. The box encloses both
the main
bending field and the entry and exit fringe fields. The straight arms at
the ends
are sufficiently long that the surface normal field and surface scalar
potential
make negligible contributions on the entry and exit faces. Also shown is a
sample
reference trajectory that, as expected, is almost straight at the ends and
bent in
the middle.

\section{Conclusions}
A new method has been developed for the computation of charged-particle
transfer maps
for general fields and geometries based on the use of surface
(boundary-value)
data.  The method requires a knowledge of all three field components on the
surface
(or, equivalently, the value of the normal field component and the scalar
potential
on the surface).  These surface values are convolved with explicitly known
and
geometry-independent kernels to produce interior fields.  The kernels
themselves are
obtained by the use of Helmholtz's theorem and Dirac magnetic monopole
vector
potentials.  The resulting interior fields satisfy the Maxwell equations
exactly and
are analytic functions of position even if the surface data contains errors
and/or the convolutions are only performed approximately.  Thus, the
resulting
transfer maps are expected to be optimally robust against computational
and/or
measurement errors.  Using these methods, modules can be added to existing
numerical
electromagnetic field-solving codes that would produce reliably, when
requested,
associated transfer maps to any order for arbitrary static charged-particle
beamline
elements.

\section{References}
\parindent=0in{1.  A. J. Dragt, ``Lie Methods for Nonlinear Dynamics with Applications to
Accelerator Physics'', University of Maryland Physics Department technical
report, (2000), Section 1.6.

2. {\it The Art and Science of Magnet Design: Selected Notes
of Klaus
Halbach}, LBL Pub-755, Lawrence Berkeley Laboratory, Feb. 1995, p. 65.

3. M. Venturini and A. Dragt, ``Accurate Computation of Transfer Maps from
Magnetic Field Data'', Nucl. Inst. and Meth. A 427, pp. 387-392 (1999).

4 A. J. Dragt et al., MaryLie 3.0 amd 5.0 manuals, University of Maryland
Physics Department technical reports, (2000).

5. P. L. Walstrom, private communication.

6. H. Wind, ``Evaluating a Magnetic Field Component from Boundary
Observations Only'', Nucl. Inst. and Meth. 84 (1970) pp. 117-124.

7. B. Hartmann, M. Berz, and H. Wollnik, ``The Computation of Aberrations
of Fringing Fields of Magnetic Multipole and Sector Magnets Using
Differential Algebra'', Nuc. Inst. and Meth. A297 (1990) pp. 343-353.

8. M. Berz, {\it Modern Map Methods in Particle Beam Physics}, Vol. 108 of 
{\it Advances in Imaging and Electron Physics},
Academic Press, San Diego, 1999, pp. 132-144.

9. R. Plonsey, R.E. Collin, {\it Principles and Applications of
Electromagnetic Fields}, The Maple Press Company, PA, 1961.

10. J. D. Jackson, {\it Classical Electrodynamics}, 3rd Ed., John Wiley and
Sons, N.Y., 1999, p. 278. }

11. J. Schwinger, L. DeRaad, et al., {\it Classical Electrodynamics}, Perseus
Books, 1998, Chapters 2 and 30.

12.  Ref. 1, Chapter 26.

13. Ref. 10, p. 36, Eq. 1.36.

\vspace {.5in}

\textbf{\Large Appendix A Sources on Surfaces of Rotation in Cylindrical
Coordinates}{\Large \par}

\vspace {0.3in}

Sources on surfaces with rotational symmetry can be used to represent the
fields in
the special \parindent=12ptcase of magnetostatic or electrostatic elements
with a
straight or slightly curved reference trajectory (e.g., quadrupoles or
dipoles
with small sagitta).  A cylindrical coordinate system is the most
convenient one to
use in treating this case. Field-point coordinates in the following are
denoted by
$(r,\theta,z)$, and source coordinates by $(a,\phi,z^\prime )$. A toroidal
surface
that encloses the magnet and has a hole through which the beam pipe passes
can be
generated by rotating the appropriate closed space curve around some axis.
A special
case of this is an infinite cylinder, for which the space curve closes at
infinity. 
It should be noted that in some accelerator applications, the beam pipe
extends into
the spaces between quadrupole pole pieces and encloses points with radial
distances
from the magnet axis that are larger than the pole-tip radius. In these
cases, no
single series-expansion map can represent the entire beam-pipe volume. If
one is
interested in a map for regions beyond the pole-tip radius, it is necessary
to use
an offset reference trajectory to ensure that the volume of interest is
within the
radius of convergence of the field expansion. The general space curve can
be defined
by two functions of arc length $s$ or some other convenient curve
parameter, i.e.,
$a=A(s)$, $z=Z(s)$. The sources can be represented by Fourier series in
$\phi$ with
coefficients that are functions of arc length $s$ and the fields and
potentials by
Fourier series in
$\theta$ with coefficients that are functions of $r$ and $z$. As in the
case of a
general surface, the sources are a superposition of a fictitious single
surface-charge distribution arising from the normal component of field at
the surface
and a fictitious double-layer distribution (i.e., normally-pointing surface
dipole-density distribution) that arises from the tangential components of
the field
at the surface. In the even simpler case of an infinitely long cylinder, it
is
straightforward to find either a double-layer or single-layer distribution
that alone
represents the fields inside the cylinder (existing MARYLIE computer
subroutines
that treat this case were described in Section 3). If the surface does not
touch
real sources, distributions are continuous except at $s$ values where the
curve
tangent is not continuous. The fictitious surface charge distribution
arising from
the normal component of the field at the surface can be written in the
form
$$
{\bf B}(s,\phi)\cdot{\bf n}(s,\phi)=\sum_{m=1}^\infty f_m(s)\sin
(m\phi).\eqno(A.1)
$$
If the sources have no solenoidal component (i.e., no net current flows
completely
around the hole of the torus), $m$ takes on the values $m=1, 2, 3, \cdot
\cdot
\cdot$. The lack of terms with $m=0$ will be assumed in the following. (In
({\it
A}.1), the cosine terms are omitted for brevity. This corresponds to
admitting
only so-called normal multipole magnets. It is straightforward to add
cosine
terms if required by the symmetry of the problem). For a particular $s$
value, each
term in ({\it A}.1) corresponds to a sinusoidal charge ring of radius
$a(s)$
at axial position $z^\prime (s)$.

The fictitious double-layer (surface dipole-density) distribution ${\bf d}$
that arises from the tangential field at the surface can be written in the
form
$$
{\bf d}(s,\phi)={\bf n}(s,\phi)\psi(s,\phi). \eqno(A.2)
$$
As
in the general three-dimensional case, the double-layer density is the
magnetic
scalar potential at the surface, i.e,
$\psi(s,\phi)=\Phi\{{\bf r}^\prime[(a(s),\phi]\}$.
If the scalar potential is to be obtained from measurements, it is
suffficient,
for example, to measure $B_\phi$ everywhere on the surface, but measure the
tangential component of ${\bf B}$ perpendicular to $B_\phi$ along the
surface
for only one value of $\phi$. The fictitious dipole-density distribution
arising
from the tangential components of field at the surface can be written in
the
form
$$
\psi(s,\phi)=\sum_{m=1}^\infty h_m(s)\sin (m\phi).\eqno(A.3)
$$

The scalar potential for the $m$th term in ({\it A}.1) (the single-layer
density
distribution) is given by the following integral:
$$
\Phi_m (r,\theta,z)={{\sin m\theta}\over{2\pi}}\oint a(s)f_m
(s)ds\int_0^\pi{{\cos m\alpha}\over{R}}d\alpha,\eqno(A.4)
$$
where
$$
R=\bigl\{a(s)^2+r^2+[z-z^\prime(s)]^2-2a(s)r\cos \alpha\bigr\}^{1/2}.
\eqno(A.5)
$$
The integral over $\alpha$ in ({\it A}.4) can be expressed in terms of a
Legendre function of the second kind,
$$
\int_0^\pi{{\cos m\alpha}\over{R}}d\alpha = {1\over{ \sqrt{a(s)r} } } Q_{m-1/2}\Bigl({{a(s)^2+r^2+[z-z^\prime
(s)]^2}\over{2a(s)r}}\Bigr)
.\eqno
(A.6)
$$
Eq. ({\it A}.4) then becomes
$$
\Phi_m(r,\theta,z)={{\sin m\theta}\over{2\pi}}\oint a(s)f_m
(s){1\over{ \sqrt{a(s)r} } } Q_{m-1/2}\Bigl({{a(s)^2+r^2+[z-z^\prime
(s)]^2}\over{2a(s)r}}\Bigr) ds.\eqno(A.7)
$$
The function of ({\it A}.6) times the $\sin m\theta$ factor is the
three-dimensional cylindrical analog of the line potential of two-dimension
potential
theory and is also logarithmically divergent as the field point approaches
the
source.  It is the scalar potential of a ring of radius $a(s)$ with a line
charge
density that varies as $\sin m\phi$. For points $r,z$ close to
$a,z^\prime$, the
potential varies as the logarithm of the distance times $\sin m\theta$.

For Hamiltonian dynamics calculations, a vector potential that gives the
same
fields as the scalar potential of ({\it A}.7) is desired. The vector
potential for
the $m$th term in ({\it A}.1) can be found by integrating the Dirac
monopole
expressions over the surface. However, simpler expressions for the vector
potential
can be obtained (this amounts to a gauge transformation of the Dirac
monopole
expressions) directly from the scalar potential in a gauge with $A^\theta
=0$
by equating the magnetic field components from the gradient of ({\it A}.7)
to the
field components from the curl of the vector potential. Simple integration
over
$\theta$ gives the two vector-potential components $A^r$ and $A^z$:

$$
A_m^r(r,\theta,z)=-{{r\cos m\theta }\over{2m\pi}}\oint a(s)f_m
(s) {{\partial}\over{\partial z}}\Bigl[ {1\over{ \sqrt{a(s)r} } } Q_{m-1/2}\Bigl({{a(s)^2+r^2+[z-z^\prime
(s)]^2}\over{2a(s)r}}\Bigr)\Bigr]ds,\eqno(A.8)
$$
and
$$
A_m^z (r,\theta,z)={{r\cos m\theta }\over{2m\pi}}\oint a(s)f_m (s)
{{\partial}\over{\partial r}}\Bigl[ {1\over{ \sqrt{a(s)r} } } Q_{m-1/2}\Bigl({{a(s)^2+r^2+[z-z^\prime
(s)]^2}\over{2a(s)r}}\Bigr)\Bigr]ds.\eqno(A.9)
$$
The derivatives in the above expressions can be evaluated explicitly using
the
derivative relations for Legendre functions, and the chain rule. The
Legendre
functions themselves can be computed for large arguments by recursion with
the
complete elliptic integrals $K$ and $E$ with the appropriate arguments, and
for
small arguments from the hypergeometric series expressions$^1$ . The scalar
potential for the $m$th Fourier component of the fictitious surface
dipole-density
sources arising from the tangential field components at the surface (see
({\it
A}.3)) can also be expressed in terms of Legendre functions:
$$
\Phi_m(r,\theta,z)={{\sin m\theta }\over{2\pi}}\oint a(s)h_m (s) {\bf n}
\cdot
\Bigl(\hat{a}{{\partial}\over{\partial a}}-\hat{z}{{\partial}\over{\partial
z}}\Bigr)\Bigl[ {1\over{ \sqrt{a(s)r} } } Q_{m-1/2}\Bigl({{a(s)^2+r^2+[z-z^\prime
(s)]^2}\over{2a(s)r}}\Bigr)\Bigr]ds.\eqno(A.10)
$$

In ({\it A}.10) , $\hat{a}$ is a radial unit vector and $\hat{z}$ an axial
unit
vector. The vector-potential components are again obtained in a gauge with
$A^\theta =0$ by equating the gradient of the scalar potential to the curl
of the
vector potential, and integrating over angle. The result is
$$
A_m^r(r,\theta,z)=-{{r\cos m\theta }\over{2m\pi}}\oint a(s)h_m (s) {\bf n}
\cdot\Bigl(\hat{a}{{\partial^2}\over{\partial a\partial z}} -
\hat{z}{{\partial^2}\over{\partial z^2}}\Bigr)\Bigl[
{1\over{ \sqrt{a(s)r} } } Q_{m-1/2}\Bigl({{a(s)^2+r^2+[z-z^\prime
(s)]^2}\over{2a(s)r}}\Bigr)\Bigr]ds,\eqno(A.11)
$$
and
$$
A_m^z(r,\theta,z)={{r\cos m\theta
}\over{2m\pi}}\oint a(s)h_m (s) {\bf n}
\cdot\Bigl(\hat{a}{{\partial^2}\over{\partial
a\partial r}}-\hat{z}{{\partial^2}\over{\partial z\partial r}}\Bigr)\Bigl[
{1\over{ \sqrt{a(s)r} } } Q_{m-1/2}\Bigl({{a(s)^2+r^2+[z-z^\prime
(s)]^2}\over{2a(s)r}}\Bigr)\Bigr]ds.\eqno(A.12)
$$
For series expansion of the potentials and fields around the axis, which is
needed
for numerical map generation, Dougall's integral expression$^2$ with the
on-axis
generalized gradient is used to extend the potential to non-zero values of
$r$. Given
the $m$th Fourier component of the scalar potential $V_m$ in cylindrical 
coordinates, we first note that the leading behavior of $V_m(r,z)$ with $r$ near
the axis
is $V_m\sim r^m$, and then define the on-axis generalized gradient
to be
$$
g_m(z)=\lim_{r\rightarrow 0}{{mV_m(r,z)}\over{r^m}}.\eqno(A.13)
$$
(In {\it A}.13, the factor of $m$ is included to make $g_m$ consistent with
the
usual definition of quadrupole gradient for $m=2$). Dougall's integral
expression
is
$$
V_m(r,z)={{(m-1)!r^m}\over{\Gamma(m+1/2)\Gamma(1/2)}}\int_0^\pi g_m(z +
ir\cos
t)\sin^{2m}tdt.\eqno(A.14)
$$
The series expansion for $V_m$ near the axis is obtained by replacing $g_m
(z +
ir\cos t)$ by its Taylor expansion, with $ir\cos t$ as the expansion
parameter, and
integrating term by term. The imaginary terms integrate to zero,
leaving the
well-known series expansion
$$
V_m(r,z)=(m-1)!r^m\sum_{n=0}^\infty
{{1}\over{n!(n+m)!}}\Bigl(-{{r^2}\over{4}}{{d^2}\over{dz^2}}\Bigr)^ng_m(z)\eqno(A.15)
$$
For the particular case of Legendre-function expressions of ({\it
A}.6)-({\it A}.12),
the hypergeometric- function representation for $Q_{m-1/2}$ is used to find
their
limiting behavior with $r$ near the axis [i.e., evaluate the limit of ({\it
A}.13)].
The series expansion for the Legendre-function factor ({\it A}.6) is then
found to
be
$$
{1\over{ \sqrt{a(s)r} } } Q_{m-1/2}\Bigl({{a(s)^2+r^2+[z-z^\prime
(s)]^2}\over{2a(s)r}}\Bigr) =
$$
$$
{{\pi a^m r^m (2m-1)!!}\over{2^m}} \sum_{n = 0}^\infty
{{1}\over{n!(n+m)!}}\Bigl(-{{r^2}\over{4}}{{d^2}\over{dz^2}}\Bigr)^n{{1}\over
{\bigl[a^2+[z-z^\prime(s)]^2\bigr]}^{m+1/2}}.
\eqno(A.16)
$$
The series expansions for the scalar and vector potentials are then
obtained by
substituting the right-hand side of ({\it A}.16) for the quantity
$$
{1\over{ \sqrt{a(s)r} } } Q_{m-1/2}\Bigl({{a(s)^2+r^2+[z-z^\prime
(s)]^2}\over{2a(s)r}}\Bigr) \eqno (A.17)
$$
everywhere it occurs in ({\it A}.7)-({\it A}.12). For numerical evaluation
of the
series coefficients as a function of $z$, in general it is necessary to
first obtain
the Fourier coefficients for the scalar and dipole densities as a function
of $s$ and
then perform the integrals inside the summation sign over $s$ by numerical
quadrature. It is possible to perform the integrals analytically if Fourier
coefficients of the source densities are represented as
piecewise-continuous
polynomials in $s$.

Finally, in reference to the problem of fitting three-dimensional field
data  on or
near the axis in straight-axis systems and extending it outward in radius
(a
mathematically unstable procedure, which the surface-data method of this
paper
avoids), one could first Fourier-analyze the data in azimuthal angle,
and treat each Fourier component as a separate
fitting
problem in $r$ and $z$ using the appropriate derivatives of the $m$th 
component of ($A$.15) times 
sin $m\theta$ or cos $m\theta$, together with a trial function for
$g_m(z)$.  One might, for example, represent $g_m(z)$ by
high-order (5th
or higher) splines and find the spline coefficients by fitting fields from
({\it
A}.15) to field values specified on a surface or over a volume.  The series
of ({\it
A}.15) will end at a finite value of $n$ with this representation of $g_m$
because
splines are made up of piecewise-continuous polynomials of finite order,
while the
``true'' $g_m$ is infinitely differentiable.  This
representation
therefore can model only the lowest-order behavior of the field, for points
at relatively
small distances from the axis relative to the physical aperture of the
element. Other
methods, in which transcendental functions are used to represent $g_m$,
give a series
that does not truncate, but the improvement can be illusory because the
higher
derivatives and the associated higher-order terms can be wrong, and in many
cases,
wildly wrong. The fundamental problem is that the fitting function that
represents
$g_m$ is generally not the ``true'' function and if it has
simple
poles, the poles, which represent sinusoidal ring sources, have little to
do with the
``true'' sources.

\vspace{0.5in}\parindent=0pt\textbf{\Large References (Appendix A)}
\vspace{0.3in}

1. M. Abramowitz and I. Stegun, Ed., {\it Handbook of Mathematical
Functions},
Dover Publications, NY , 1972 (a reprint of the NBS \#55, Applied
Mathematics
Series of 1964 with the same title), pp. 331-337.

2. E. T. Whittaker and G. N. Watson, {\it A Course of Modern Analysis},
Fourth
Edition (reprinted), Cambridge University Press, 1963, p. 400.

\vspace{0.5in}

\textbf{\Large Appendix B Surface Topologies and Dirac Monopole
Orientation}
\vspace{0.3in}

Various surfaces in addition to the curved box and infinite cylinder
already
discussed can be considered. For example, toroidal surfaces of the type
shown
in Fig. B.1 can be made to have minimal clearance of the magnetic element
on
the ends and have a hole that is large enough to accomodate the beam pipe. 
Since
the beam goes through the hole, and not through the surface, the two ends
of the
surface do not need to extend out to field-free regions and the surface can
be
shorter. Also, if the vertical surfaces at the ends are made to extend far
enough out
to the sides that the fields on them become negligible at the outer edges,
source
strengths on the outside surfaces connecting them are negligible.  In such
a case
source distributions need only be found for inner surface around the hole
and for the
parts of the vertical end surfaces near the ends of the hole. In most
cases, all of
the pieces of the surfaces can be constructed from flat pieces and parts of
cylindrical surfaces. For example, for curved dipoles, the vertical parts
of the
inner surface could be made of segments of cylinders.

\vspace{0.3cm}
{\par\centering \includegraphics{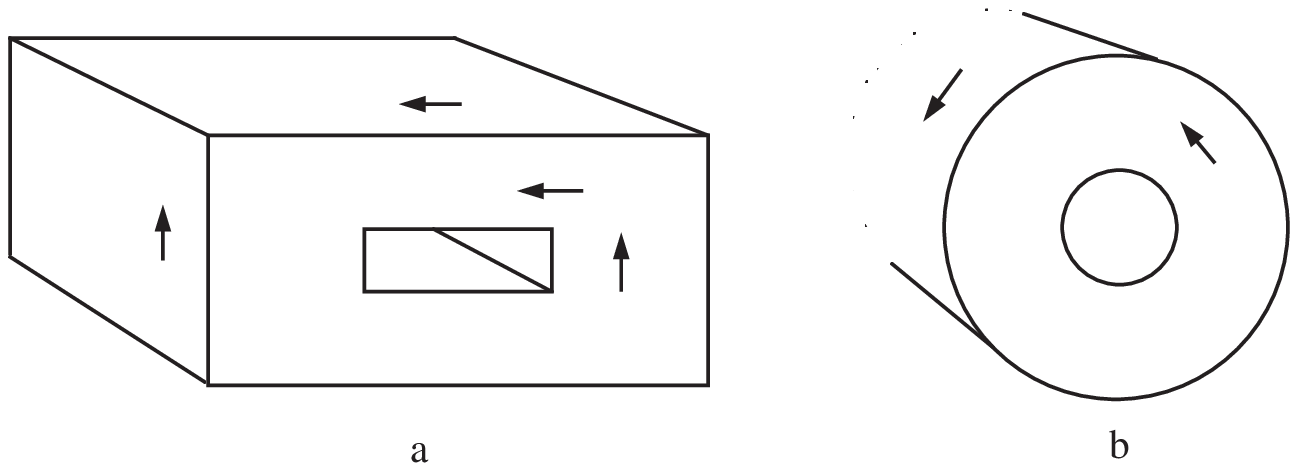} \par}
\vspace{0.3cm}

Fig. B.1 Two possible toroidal surfaces with Dirac string orientations
indicated.
Surface a would be used with an H magnet, while b would be used with a
quadrupole.

\parindent=12pt
\vspace{0.3cm}

It should be noted that the discontinuities in the tangents to the surface
are not a
problem in principle because the surfaces are fictitious and have vacuum on
each
side; the is no field peaking as there is at the edge of a magnet yoke,
etc. However,
attention must be paid to numerical problems related to discrete source
placement and
spacing along edges.

It can be seen that if toroidal surfaces are used, use of Dirac monopoles
with
strings perpendicular to the end surfaces could be a problem in curved
beamlines, as
they could intersect the beam volume. This problem can be avoided by using
strings
tangential to the surface as shown in Fig. B.1. If tangential strings are
used, a
two-sided Dirac monopole with two half-strength strings going in opposite
directions
can be used; the associated expression for the vector potential is somewhat
simpler
and may have some advantages in computation: 
$$
{\bf G}^n({\bf r}; {\bf r}^{\prime}, {\bf m})={{{\bf m} \times {({\bf
r}-{\bf
r}^{\prime})}} \over {4\pi}|{\bf m} \times {({\bf r}-{\bf
r}^{\prime})|^2}}.\eqno
(B.1)
$$

\end{document}